\documentclass[pre,aps,twocolumn]{revtex4-2}
\usepackage{amsfonts}
\usepackage{times}
\usepackage{amsmath, amsthm}
\usepackage[colorlinks=true,citecolor=blue]{hyperref}
\usepackage[final]{graphicx}
\usepackage{amssymb, algorithm, algpseudocode}
\algnewcommand\algorithmicforeach{\textbf{for each}}
\algdef{S}[FOR]{ForEach}[1]{\algorithmicforeach\ #1\ \algorithmicdo}
\usepackage[title]{appendix}
\usepackage{xcolor,comment}

\begin{document}

\title{On the quasi-continuum approximation of
some  
localized patterns in the
FPUT lattice}

\author{Su Yang}

\affiliation{Department of Mathematics and Statistics, University
of Massachusetts Amherst, Massachusetts 01003-4515, USA}

\author{Wenrong Sun}
\affiliation{School of Mathematics and Physics, University of Science
and Technology Beijing, Beijing 100083, China}

\author{Lei Liu}
\affiliation{College of Mathematics and Statistics, Key Laboratory of Nonlinear Analysis and its Applications, Ministry of Education, Chongqing University, Chongqing 401331, China}

\author{P. G. Kevrekidis}
\affiliation{Department of Mathematics and Statistics, University
of Massachusetts Amherst, Massachusetts 01003-4515, USA}
\affiliation{Department of Physics, University of Massachusetts Amherst, Massachusetts 01003-4515,USA}
\affiliation{Department of Mechanical Engineering, Seoul National University,
1 Gwanak-ro, Gwanak-gu, Seoul 08826, South Korea}

\date{\small\today}

\begin{abstract}
    In the present work, we present a number of localized wave patterns that are theoretically analyzed and
    numerically illustrated to be observable  within the 
    widely applicable paradigm of the FPUT lattice. In particular, we derive a modified KdV equation from the FPUT lattice, which admits a variety of localized waves including these exact rational solutions representing rogue-wave profiles, solitons and breathers on the top of not only homogeneous,
    but also periodic elliptic function traveling-wave background. We utilize these exact solutions of the modified KdV reduction to construct consistent initial conditions for the FPUT lattice and perform time stepping of the latter. Relevant comparisons between these numerical solutions of the FPUT lattice and their associated analytical counterparts have been conducted to demonstrate good performance of the derived modified KdV reduction in approximating distinct localized wave structures from the FPUT lattice.
    This approach paves the way for importing a number of 
    quasi-continuum waveforms to the FPUT lattice and the
    potential associated physical experiments, including recent
    ones in mechanical metamaterials.
\end{abstract}

\maketitle

\section{Introduction}

Nonlinear wave phenomena are ubiquitous in a variety of mathematical models and physical systems~\cite{Ablowitz2011a}. Typical wave patterns include solitons \cite{ablowitz1981solitons,newell1985solitons,carretero2024nonlinear}, dispersive shock waves \cite{EL201611} and rogue waves \cite{KharifPelinovskySlunyaev2009,DudleyDiasErkintaloGenty2014}. These wave structures can be numerically observed, for instance, in the Fermi–Pasta–Ulam–Tsingou (FPUT) lattices \cite{FPUreview,VAINCHTEIN2022}, nonlinear Schr\"odinger (NLS) models \cite{AblowitzPrinariTrubatch,Kevrekidis2009}. Among these 
patterns, recently, rogue waves, also known as freak waves, have drawn considerable interest  not only due to their numerical existence, but also its emergence in physical experimental settings including nonlinear optics \cite{kibler2010peregrine,solli2007optical}, plasma physics \cite{bailung2011observation}, and hydrodynamics \cite{chabchoub2011rogue,chabchoub2012super}, as well as 
most recently in mechanical metamaterial lattices
described by FPUT type systems~\cite{miyazawa2026formationmechanicalroguewaves}.

The study of extreme events at the mathematical level can
be traced back to nearly half a century and the seminal work of Peregrine \cite{peregrine1983water}, where the exact Peregrine soliton solution was rigorously derived for the NLS equation to model the spatio-temporal profiles of the rogue waves. Thereafter, extensive investigations have taken place, especially so at the level
of NLS-type equations~\cite{DudleyDiasErkintaloGenty2014,OnoratoResitoriBaronio2016,AkhmedievPelinovsky2010} which continue to this day,
through the study of high- (and even infinite-) order
rogue waves~\cite{He_2013,suleimanov,Bilman_2020}.
It is interesting to mention,  in the context of
the lattice model level of interest herein, that such
rogue waves have been proposed to be possible to adapt
to granular crystals and related metamaterial
systems not only at the level of single-component
lattice systems~\cite{PhysRevE.98.032903}, but also
at the one of multi-component systems combining, e.g.,
translational and rotational degrees of freedom as in the
case of~\cite{miyazawa2}.

A separate thread of related activity has taken place 
at the level of
the modified KdV (mKdV) equation which assumes the following standard form
\begin{equation}\label{standard mKdV}
    u_t + \sigma u^2u_{x} + \gamma u_{xxx} = 0.
\end{equation}
Here $\sigma, \gamma \in \mathbb{R}$ refer to the coefficients of the nonlinearity and dispersion. Eq.~\eqref{standard mKdV} admits numerous interesting exact solutions including the multi-order rational solutions \cite{PhysRevE.99.050201, ankiewicz2018rogue}, breather \cite{clarke2000generation,kruglov2021breather} and kink solutions \cite{el2017dispersive}. 
The mKdV equation \eqref{standard mKdV} possesses numerous solutions representing intriguing wave patterns due to the non-convexity property of the nonlinearity. Using Darboux transformations \cite{MatveevSalle1991,mucalica2022solitons}, further analytical solutions can be found such as the soliton solution placed on the dnoidal wave background, as illustrated in the work  of~\cite{chen2018rogue}. The overarching goal of the present work 
parallels that of earlier studies such as~\cite{PhysRevE.98.032903,miyazawa2} and
consists of leveraging the wealth of exact solutions of the mKdV equation \eqref{standard mKdV} to explore the associated wave dynamics in the FPUT lattice. A significant advantage that we draw upon
herein is that we shall not only take advantages of the different orders of rational solutions of Eq.~\eqref{standard mKdV}, but 
we will explore 
localized waves on top of periodic traveling backgrounds. While optics and hydrodynamics have
served as settings where such rogue patterns have emerged
on top of periodic backgrounds, the recent work of~\cite{miyazawa2026formationmechanicalroguewaves} renders
especially relevant and timely their potential appearance
in mechanical metamaterials, thus motivating the present study.

Our presentation is structured as follows. In Sec.~\ref{Sec: Intro and models}, we introduce the FPUT lattice which shall be the main model of interest throughout this work, and simultaneously elaborate on the motivation toward deriving and applying the modified KdV reduction to approximate different localized wave patterns in the FPUT lattice. Next, in Sec.~\ref{Sec: Local waves on constant back}, we revisit some important exact solutions of the derived mKdV reduction, which are located on top of homogeneous backgrounds and serve to model rogue-wave features. We then use these exact solutions to set up proper initial conditions for the FPUT lattice and perform time evolution through suitable numerical solvers of the nonlinear dynamical lattice. Numerical comparisons have been conducted at distinct time snapshots to probe the performance of these analytical exact solutions of the mKdV reduction toward successfully approximating the associated dynamics of the FPUT lattice. In Sec.~\ref{Sec: Waves on periodic backgrounds}, we turn our attention to solutions placed on the periodic-traveling wave backgrounds and also establish analogous numerical comparisons as in Sec.~\ref{Sec: Local waves on constant back}. Finally, this work ends within Sec.~\ref{Sec: conclusions and open directions} with an overall summary of the findings and some interesting open directions for future studies.

\section{Model descriptions and theoretical setup}\label{Sec: Intro and models}

In this paper, our principal focus is
on the investigation of the following FPUT lattice~\cite{FPUreview,VAINCHTEIN2022}:
\begin{equation}\label{FPUT lattice in displacement}
    \frac{d^2u_n}{dt^2} = V'(u_{n+1}-u_n) - V'(u_n-u_{n-1}),
\end{equation}
with
\begin{equation}
    V'(x) = K_2x + K_3x^2 + K_4x^3,
\end{equation}
where $n \in \mathbb{Z}$, $K_{2,3,4} \in \mathbb{R}$ are three constants, and $u_n(t) \in \mathbb{R}$ represents the displacement of the $n$th particle from its equilibrium position at time $t$. 

Now, we introduce the strain variable defined as $y_n(t) \equiv u_{n+1}(t) - u_n(t)$ so that Eq.~\eqref{FPUT lattice in displacement} can be rewritten as follows:
\begin{equation}\label{FPUT in strain}
   \begin{aligned}
    &\frac{d^2y_n}{dt^2} = K_2\left(y_{n+1} - 2y_n + y_{n-1}\right) + K_3\left(y_{n+1}^2 - 2y_n^2 + y_{n-1}^2\right) \\
    &+ K_4\left(y_{n+1}^3-2y_n^3 + y_{n-1}^3\right).
    \end{aligned}
\end{equation}
If we look for a plane-wave ansatz in the form of an infinitesimal perturbation near the mean of $\overline{y}$: $y_n(t) = \overline{y} + \delta\exp\left[i(kn - \omega t)\right]$, where $0<\delta\ll 1$, then a direct substitution of this plane-wave ansatz into Eq.~\eqref{FPUT in strain} yields the following linear dispersion relation, upon the elimination of the smallness parameter of $\delta$,
\begin{equation}\label{ldr of the FPUT lattice}
    \omega^2 = 4\left(K_2 + 2K_3\overline{y}+3K_4\overline{y}^2\right)\sin^2\left(\frac{k}{2}\right).
\end{equation}
Without loss of generality, we consider only the right-propagating waves and take the positive branch of the linear dispersion relation in Eq.~\eqref{ldr of the FPUT lattice}, and compute the long-wave expansion of the dispersion relation to obtain that
\begin{equation}\label{positive branch ldr}
    \omega_+ = 2\sqrt{K_2+2K_3\overline{y} + 3K_4\overline{y}^2}\left(\frac{k
    }{2} - \frac{1}{48}k^3 + \mathcal{O}(k^5)\right).
\end{equation}
Ignoring the terms on the order of $\mathcal{O}(k^5)$ and setting $K_3 = 0$, i.e., focusing on the $\beta$-FPUT model~\cite{FPUreview,VAINCHTEIN2022} in what follows, yields the linear dispersion relation associated with the modified KdV equation describing small amplitude wave propagation.

In order to derive the associated mKdV reduction of Eq.~\eqref{FPUT in strain}, we assume the following multi-scale change of variables:
\begin{equation}\label{multi-scale variables}
    \begin{aligned}
        &y_n(t) = \epsilon Y(X,T),\\
        &X=\epsilon(n-c_st), \quad T = \epsilon^3 t,
    \end{aligned}
\end{equation}
where $c_s = \sqrt{K_2}$ refers to the speed of sound, and $0 < \epsilon \ll 1$ is a formal smallness parameter.

Substituting Eq.~\eqref{multi-scale variables} into Eq.~\eqref{FPUT in strain}, assuming $K_3=0$ as indicated above, and collecting relevant terms on distinct orders of $\epsilon$ yields the following
mKdV equation at $\mathcal{O}(\epsilon^5)$:
\begin{equation}\label{asympototic results}
%
         \quad Y_T + \frac{3K_4}{2c_s}Y^2Y_X + \frac{K_2}{24c_s}Y_{XXX} = 0.
\end{equation}
We notice that due to the indefinite signs of the two parameters of $K_{2,4}$, we can either have the focusing or defocusing mKdV reduction. 
Having now derived the relevant mKdV model,
we explore how to connect its potential spatio-temporally
localized waveforms with suitable lattice profiles
(and associated initializations in order to observe them).

\section{Numerical scheme and initial conditions}\label{sec: numerical setup and ICs}

We now discuss some necessary numerical preliminaries in order to solve the FPUT lattice in Eq.~\eqref{FPUT in strain} with appropriate initial data. Firstly, to numerically solve the FPUT lattice in Eq.~\eqref{FPUT in strain}, we reformulate it as the following first-order system:
\begin{align}
    &\frac{dy_n}{dt} = s_n,\label{Eq.1 first-order}\\
    &\frac{ds_n}{dt} = \nonumber K_2(y_{n+1}-2y_n+y_{n-1}) 
    + K_4(y_{n+1}^3-2y_n^3+y_{n-1}^3). \label{Eq.2 first-order}
\end{align}
where $s_n$ is defined as the velocity variable.
We then apply the standard RK4 time integration scheme to numerically solve the system in Eqs.~\eqref{Eq.1 first-order}. Moreover, we shall use a sufficiently wide computational domain so that the evolution dynamics never hits the boundaries.
Accordingly, the specific type of boundary conditions will not play any essential role in the numerical simulations that follow. Next, 
we assume that the exact analytical solution for the mKdV equation in Eq.~\eqref{asympototic results} reads: $Y = \widetilde{Y}(X,T)$. Then, the initial data for the strain of $y_n$ at time $t = t_0$ simply reads: $y_n(t_0) = \epsilon \widetilde{Y}(X,T_0)$ where $T_0 = \epsilon^3 t_0$. For $s_n$, we observe that, by the chain rule,
\begin{equation}\label{eq: relation to construct IC for sn}
   \begin{aligned}
    s_n(t) &= \frac{dy_n}{dt} \\
    &= \epsilon \partial_tY\\
    &= \epsilon\left(Y_{T}\frac{\partial T}{\partial t} + Y_X\frac{\partial X}{\partial t}\right)\\
    &= \epsilon^4Y_T - c_s\epsilon^2Y_X.
    \end{aligned}
\end{equation}
Hence, the initial condition for the velocity of $s_n$ at $t = t_0$ reads:
\begin{equation}\label{IC for sn}
    s_n(t_0) = \epsilon^4\widetilde{Y}_T(X,T_0) - c_s\epsilon^2\widetilde{Y}_X(X,T_0).
\end{equation}
We note in passing that a similar path was followed 
in the work of~\cite{Yang_Biondini_Chong_Kevrekidis_2025} in the
context of granular crystals.

\section{Localized waves on homogeneous backgrounds}\label{Sec: Local waves on constant back}

In this section, we shall investigate some analytical solutions of the mKdV reduction in Eq.~\eqref{asympototic results} which can potentially represent and approximate the rogue-wave profiles of the FPUT lattice \eqref{FPUT in strain}.

\subsection{Second-order rational solution}
The first order rational solution of the mKdV is merely
a Lorentzian form traveling wave. 
Hence, we start our investigation herein with the so-called second-order rational solution, denoted as $Y_2$:
\begin{equation}\label{mKdV 2-order soln}
    Y_2(X,T) =  \frac{12G_2}{D_2} + 1,
\end{equation}
with
\begin{equation}
   \begin{aligned}
    &G_2 = 3 - (6aT + bX)\left[(6aT+bX)^3 + 6(22aT + bX)\right],\\
    &D_2 = 12aT\bigg[243(2a)^4T^4(aT+bX) \\
    &+ bX(3b^4X^4-2b^2X^2+51)\\
    &+ 72a^2bXT^2(5b^2X^2-9)\\
    &+108a^3T^3(15b^2X^2-13)\\
    &+3aT(15b^4X^4-30b^2X^2+139)\bigg]\\
    &+b^6X^6+3b^4X^4+27b^2X^2+9,
    \end{aligned}
\end{equation}
where
\begin{equation}\label{nece coefficients}
    a = \frac{b^3\gamma_3}{4}, \quad b = \sqrt{\frac{-2\beta}{3\gamma_3}},
\end{equation}
and 
\begin{equation}
\beta = \frac{3K_4}{2c_s}, \quad \gamma_3 = -\frac{K_2}{24c_s}.
\end{equation}
It is worthwhile to notice that the second-order solution \eqref{mKdV 2-order soln} admits its global maximum at $(X,T) = (0,0)$, which is 5. It is a localized wave pattern on the homogeneous background of $+1$. Moreover, it is shown in Ref.~\cite{ankiewicz2018rogue} that the $j$th order rational solution of the mKdV equation \eqref{standard mKdV} admits its maximum amplitude which is $2j + 1$ at the origin.

\begin{figure*}
    \centering
    \includegraphics[width=0.325\linewidth]{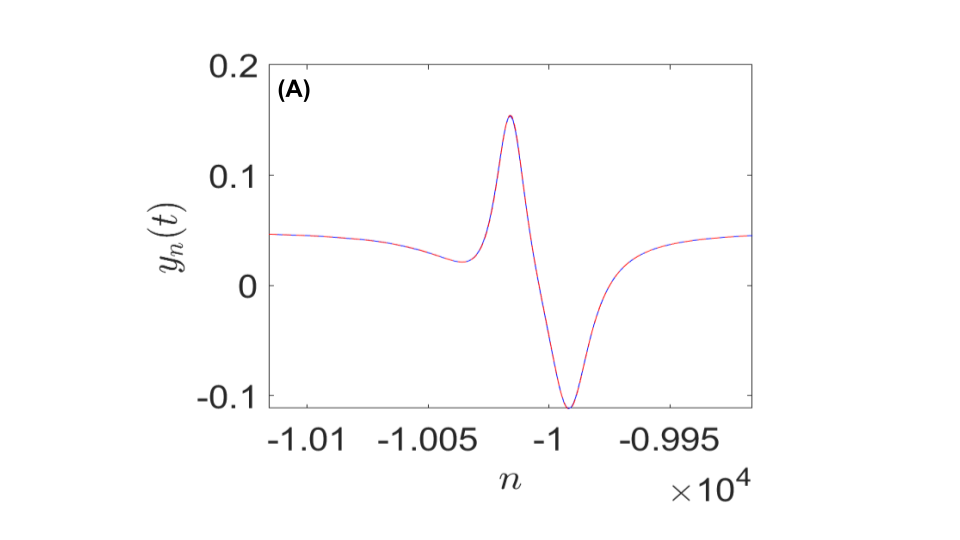}
    \hfill
    \includegraphics[width=0.325\linewidth]{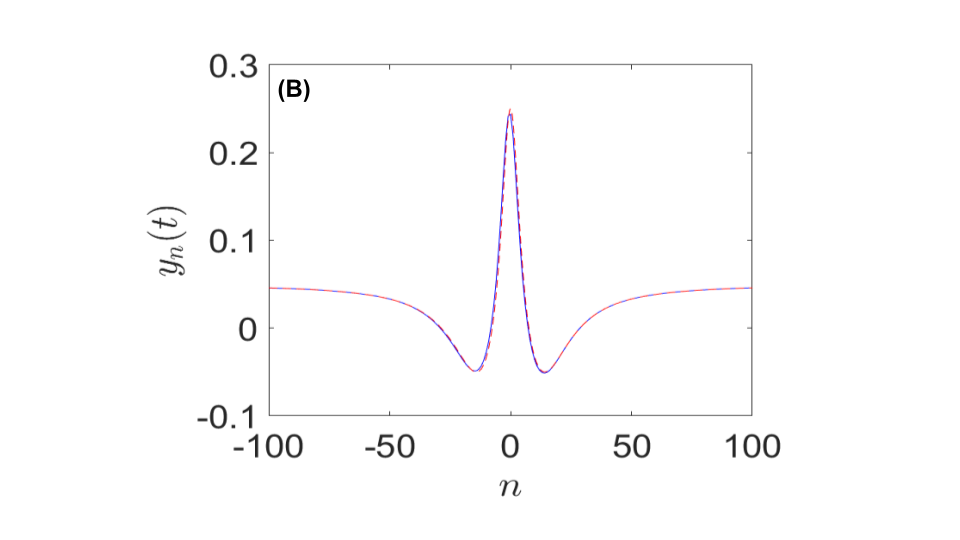}
    \hfill
    \includegraphics[width=0.325\linewidth]{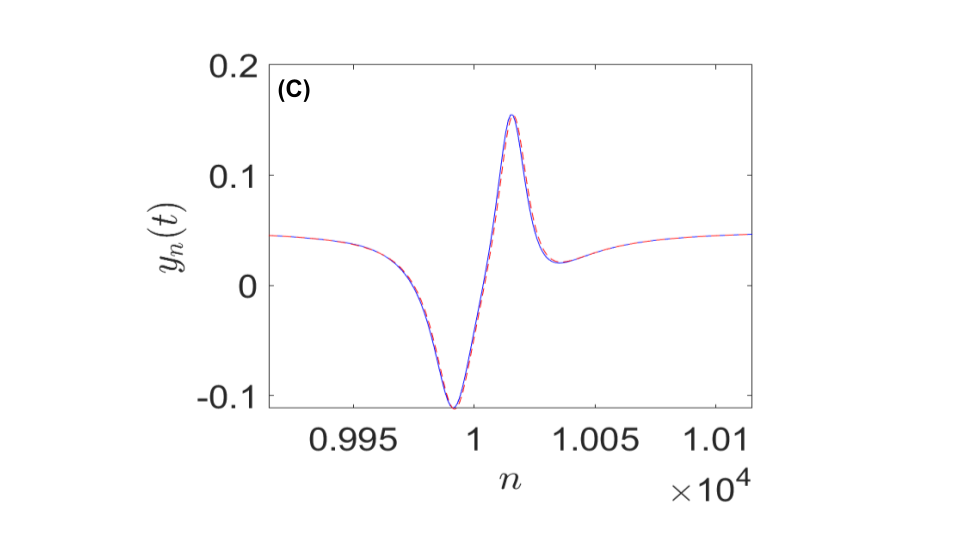}
    \hfill
    \includegraphics[width=0.325\linewidth]{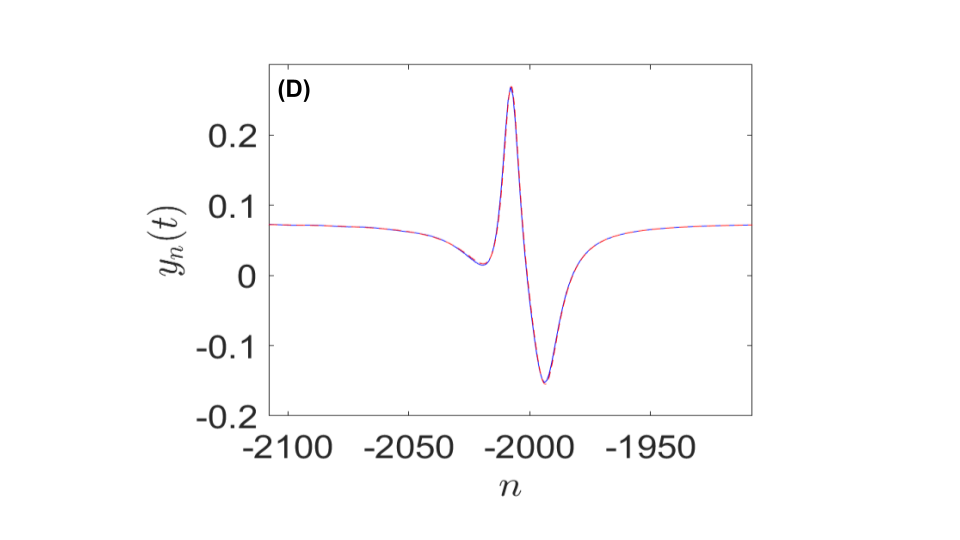}
    \hfill
    \includegraphics[width=0.325\linewidth]{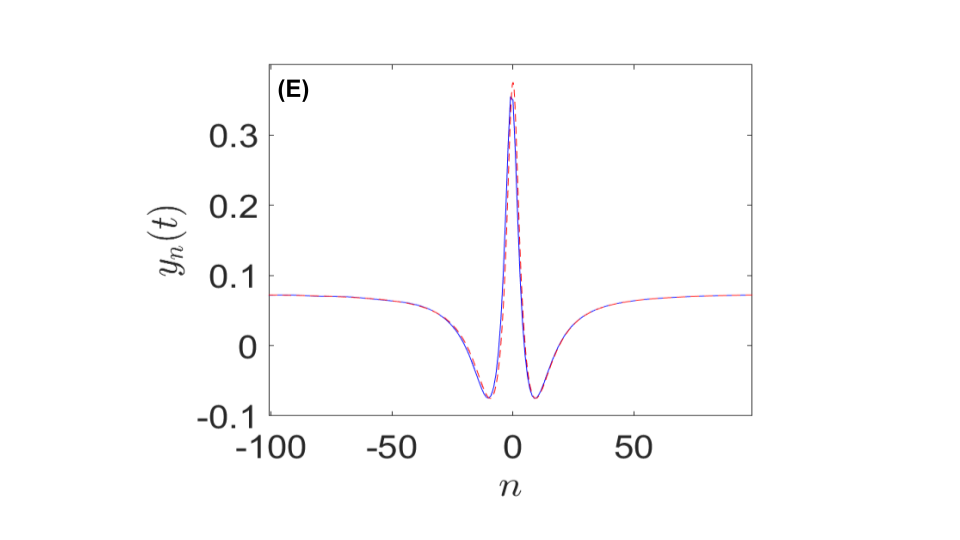}
    \hfill
    \includegraphics[width=0.325\linewidth]{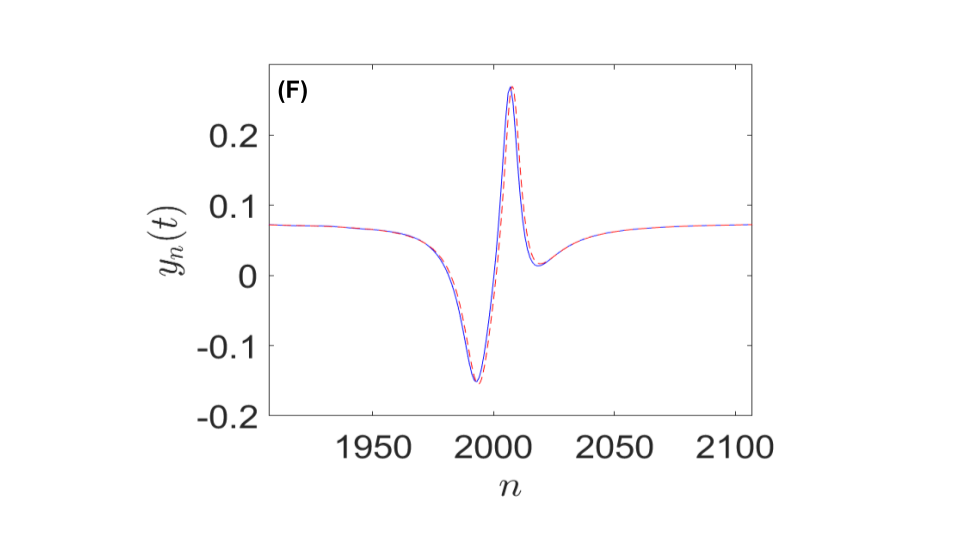}
    \caption{Comparison of the second-order rational solutions. The first and second row depict the comparisons associated with $\epsilon = 0.05, 0.075$, respectively. Note that the solid blue curves display the numerical solutions of the FPUT lattice \eqref{FPUT in strain}, and the dashed red curves showcase the exact solutions specified in Sec.~\ref{Sec: Local waves on constant back}. In addition, notice that $K_2 = 4$ and $K_4 = 1$ in all the relevant numerical simulations.}
    \label{fig:Second-order solns comparison}
\end{figure*}

Fig.~\ref{fig:Second-order solns comparison} depicts the comparisons of the numerical dynamics of the FPUT lattice \eqref{FPUT in strain} ---initialized in the way explained in the previous section--- with its associated analytical second-order rational solutions in Eq.~\eqref{mKdV 2-order soln}. In particular, for the cases of $\epsilon = 0.05, 0.075$, we initiate the numerical simulations of the FPUT with the IC specified in Sec.~\ref{sec: numerical setup and ICs} at $t_0 = -10000, -5000$, and perform time stepping up to $t = 5000, 1000$, respectively. The FPUT dynamics is compared with the Eq.~\eqref{mKdV 2-order soln} at $t = -5000, 0, 5000$ and $t = -1000, 0, 1000$ for the cases of $\epsilon = 0.05, 0.075$, and we can clearly see that despite some deviations of the maximum amplitude of the localized wave profiles, the numerical solutions fit reasonably well with their associated analytical counterparts. 
As an additional quantitative diagnostic, we compute the relative $L^{\infty}$ error measuring the deviation between the numerical exact solutions with the associated analytical solutions. Firstly, we define such error as follows:
\begin{equation}\label{eq: def of L_inf loss}
    L^{\infty}_{\epsilon}(t) = \frac{\max_{1\leq j \leq N}\left|u^{\text{ana}}_j - u^{\text{num}}_j\right|}{\max_{1\leq j \leq N}\left|u^{\text{ana}}_j - u^{\text{ana}}_{\text{back}}\right|},
\end{equation}
where $u_j^{\text{ana}} = \epsilon\widetilde{Y}\left(\epsilon(n_j-c_st),\epsilon^3t\right)$ and $u_j^{\text{num}} = y_{n_j}(t)$ denote the analytical solution of the mKdV reduction \eqref{asympototic results} and the numerical solutions of the FPUT lattice \eqref{FPUT in strain}, respectively. Moreover, $u^{\text{ana}}_{\text{back}}$ refers to the background of the solutions. Fig.~\ref{fig:L_inf loss plot} depicts the numerically computed $L^{\infty}$ losses. It is clear that the error reaches its maximum around the time when the maximal amplitude of the rational solutions is reached. This error is also higher, the higher
the order of the rational solution.
This 
qualitative and even semi-quantitative agreement between the
analytical and numerical results demonstrates the feasibility of utilizing the second-order rational solutions \eqref{mKdV 2-order soln} in order to perform a quasi-continuum approximation of a rogue-wave profile embedded in the FPUT lattice dynamics.

\subsection{Third-order rational solution}
In a similar vein to the one above, we can explore higher
order rational solutions of the mKdV~\cite{ankiewicz2018rogue}.
In particular, the mKdV equation in Eq.~\eqref{asympototic results} also admits the following third-order solution denoted as $Y_3$~\cite{ankiewicz2018rogue}:
\begin{equation}\label{third-order soln of mKdV}
    Y_3(X,T) = \frac{24G_3}{D_3} - 1,
\end{equation}
with
\begin{equation}
    \begin{aligned}
        &G_3 = 4\bigg[800c^3T^3z + 150c^2T^2\left(16z^4 - 8z^2 + 5\right)\\
        &+120cTz\left(-16z^4+40z^2+15\right) \\
        &+z^2\left[8z^2\left(32z^6+120z^4+420z^2-225\right)-675\right]\bigg] + 675,\\
        &D_3 = 2025 + 8\bigg[800c^4T^4 - 800c^3T^3z(-3+4z^2) \\
        &+30c^2T^2(165-180z^2+240z^4+64z^6)\\
        &+10cTz\left[-675+32z^2\left(45-27z^2+8z^6\right)\right]\\
        &+z^2\bigg[6075+2z^2\times\\
        &\left(3375+16z^2\left[585+135z^2+24z^4+16z^6\right]\right)\bigg]\bigg],
    \end{aligned}
\end{equation}
where $z = \frac{1}{4}(2bX - cT)$, $c = -3b^3\gamma_3$, and $b$ is given in Eq.~\eqref{nece coefficients}. It is interesting to
note that asymptotically the denominator consistently
across our examples reflects a polynomial with two powers
more in the denominator in comparison to the numerator, yet the
polynomials in each case correspond to higher orders.

\begin{figure*}
    \centering
    \includegraphics[width=0.325\linewidth]{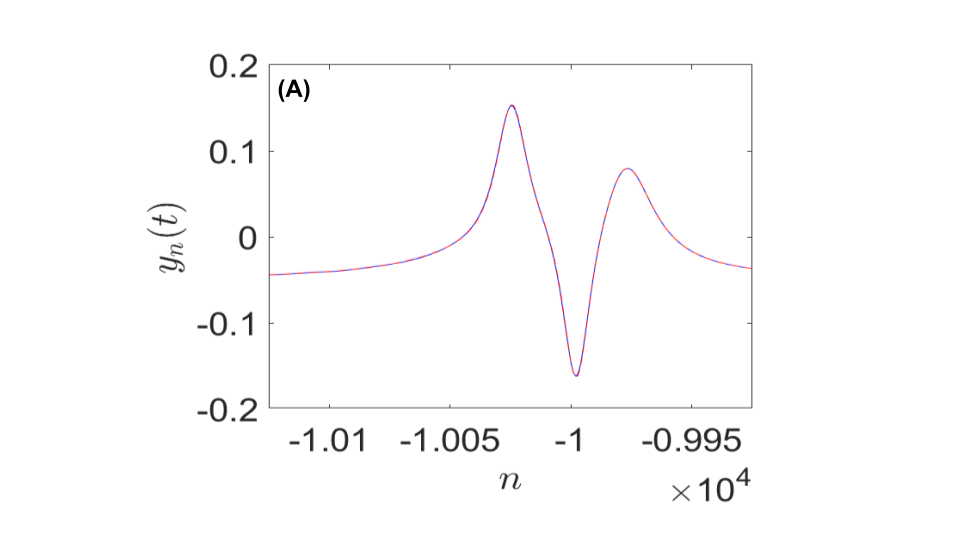}
    \hfill
    \includegraphics[width=0.325\linewidth]{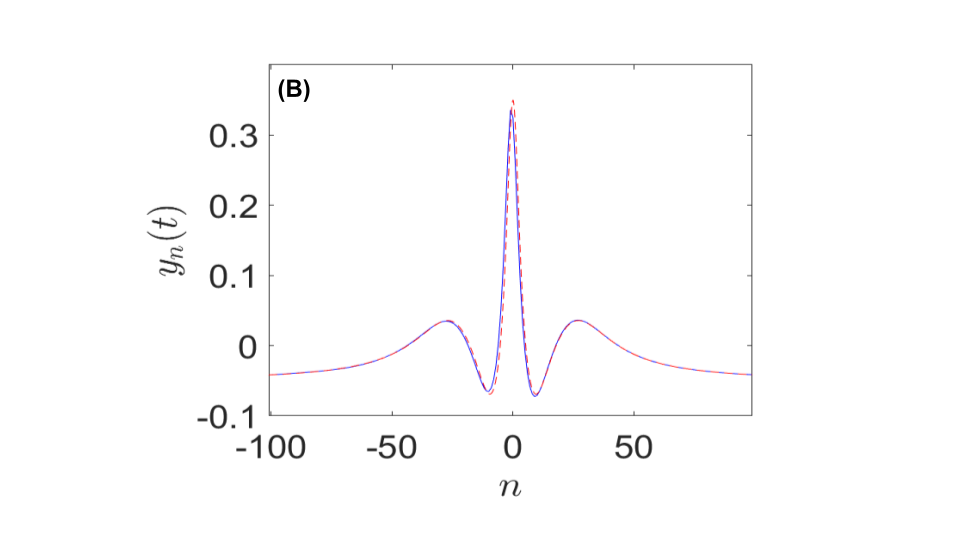}
    \hfill
    \includegraphics[width=0.325\linewidth]{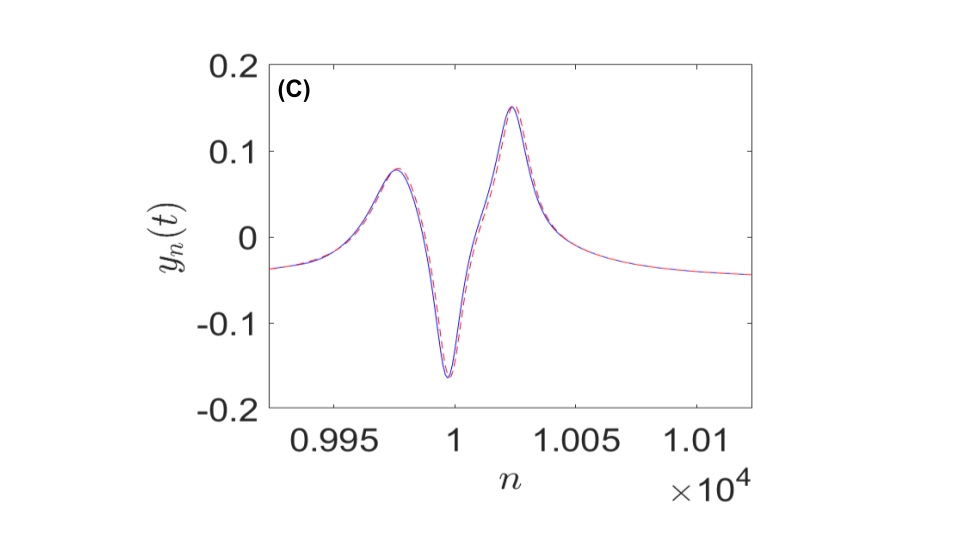}
    \hfill
    \includegraphics[width=0.325\linewidth]{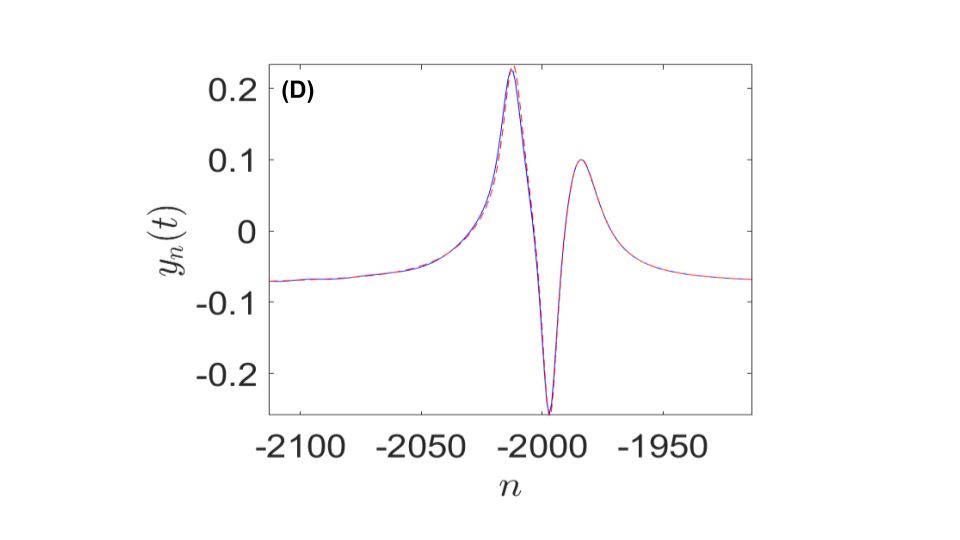}
    \hfill
    \includegraphics[width=0.325\linewidth]{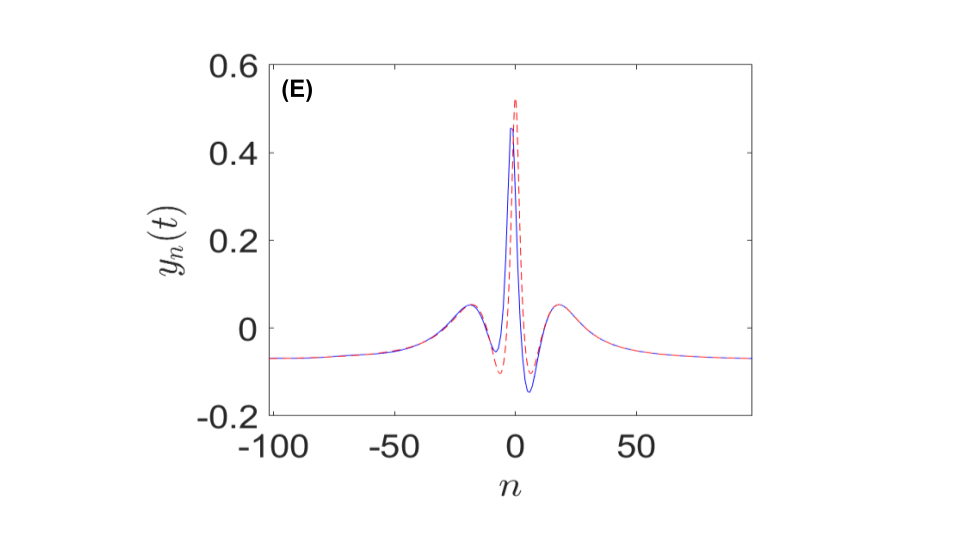}
    \hfill
    \includegraphics[width=0.325\linewidth]{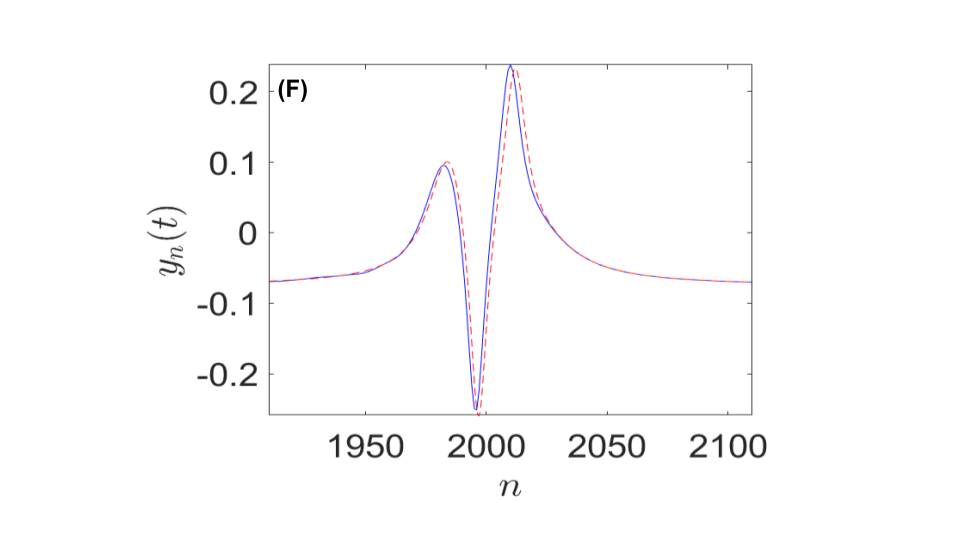}
    \caption{Same as in Fig.~\ref{fig:Second-order solns comparison} but now for the comparisons of the third-order rational solutions.}
    \label{fig:Third-order solns comparison}
\end{figure*}

\begin{figure}[b!]
    \centering
    \includegraphics[width=0.475\linewidth]{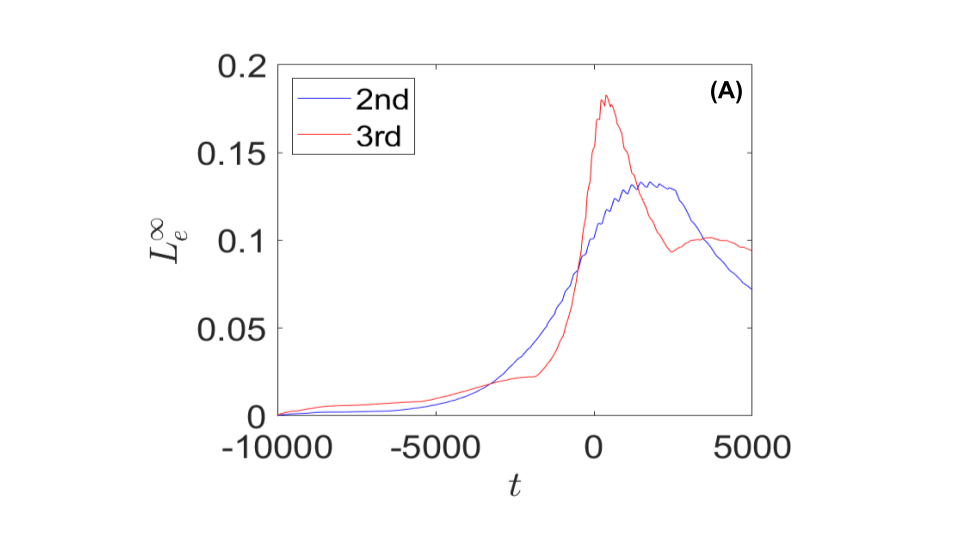}
    \hfill
    \includegraphics[width=0.475\linewidth]{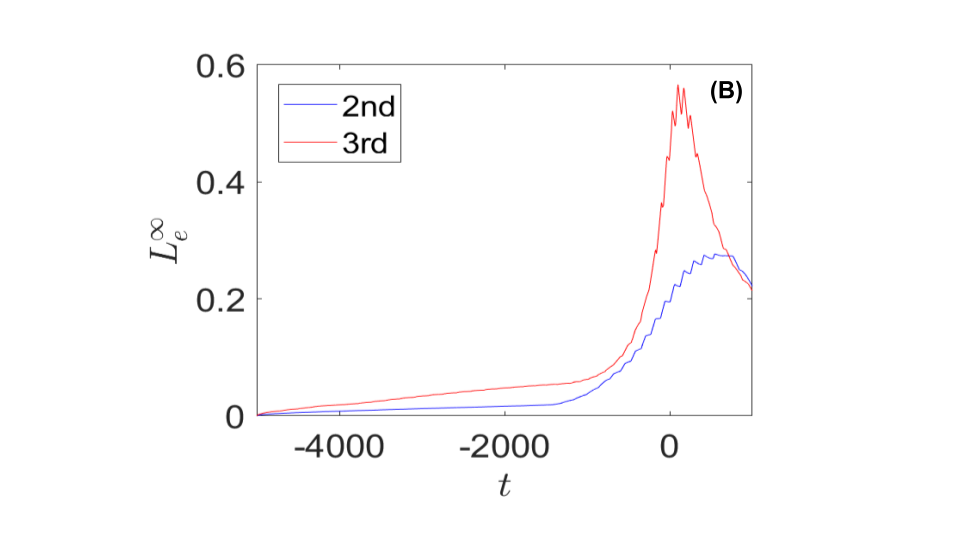}
    \caption{The relative $L^{\infty}_{e}$ losses for the $2$nd and $3$rd-order rational solutions. Panels (A) and (B) refer to the losses computed for the cases of $\epsilon = 0.05$ and $0.075$, respectively.}
    \label{fig:L_inf loss plot}
\end{figure}

We then compare the FPUT numerical dynamics where the initial condition is constructed using the third-order solution specified in Eq.~\eqref{third-order soln of mKdV} at $t_0 = -10000, -5000$ for the cases of $\epsilon = 0.05, 0.075$, respectively. Fig.~\ref{fig:Third-order solns comparison} displays these relevant comparisons where the numerical and analytical solutions are compared at $t = -5000, 0, 5000$ for $\epsilon = 0.05$, and $t = -1000, 0, 1000$ for $\epsilon = 0.075$. In this case too, we observe a very good agreement 
between the long wavelength approximation of our theory
and the numerical nonlinear dynamical (FPUT) lattice result.
Fig.~\ref{fig:L_inf loss plot} displays the $L^{\infty}$ error in this
case as well, demonstrating its increase over that of the 2nd
order solution. Yet, as the detailed comparisons of Fig.~\ref{fig:Third-order solns comparison} suggest, the relevant error stays  bounded and fairly reasonable
for sufficiently small $\epsilon$ (e.g., not exceeding 20$\%$ for
$\epsilon=0.05$).

\section{Localized waves on traveling-periodic backgrounds}\label{Sec: Waves on periodic backgrounds}

In this section, we turn our attention to the localized wave structures of the mKdV equation in Eq.~\eqref{asympototic results}
in the presence of a nontrivial (i.e., non-homogeneous)
background. More specifically, we shall use localized wave patterns which are placed on top of a periodic-traveling wave background of the mKdV reduction in Eq.~\eqref{asympototic results} in order to
generate the associated wave structures in the FPUT lattice \eqref{FPUT in strain}. 
To the best of our knowledge, in the context of FPUT lattices~\cite{VAINCHTEIN2022},
and by extension of mechanical metamaterials probed
theoretically/numerically~\cite{PhysRevE.98.032903,miyazawa2}
or experimentally~\cite{miyazawa2026formationmechanicalroguewaves},
such waveforms have not been previously proposed or 
probed in dynamical evolution. Hence, the present exploration
could form a basis not only for further theoretical/numerical
exploration, but also for lattice experimental realization
of such waveforms.

Nevertheless, it is important to mention a key caveat here.
Unlike the comparisons conducted for the rational solutions specified in Sec.~\ref{Sec: Local waves on constant back}, we shall not compare the dynamics of the FPUT lattice with the corresponding exact solutions at long-time dynamics (i.e. $\left|t\right| \gg1$). This is due to the inevitable boundary effects. In particular, when the initial condition is given by a localized wave pattern on
top of a periodic background, any type of boundary conditions (e.g. periodic boundary conditions) shall not be fully satisfied 
(at all times) due to the discreteness of the lattice model \eqref{FPUT in strain}. These undesirable boundary effects can play a detrimental role on the actual evolution dynamics of the FPUT lattice \eqref{FPUT in strain}. Hence, one has to use a considerably wide computational domain to avoid these boundary effects influencing the actual dynamics of the model, so the computational cost shall increase drastically, which causes technical issues in the numerical experiments. Therefore, accordingly, and aiming to visualize
the dynamics of the ``practically infinite'' lattice, we only compare dynamics at relevantly small $t$'s for solutions on a periodic background, and this is a nontrivial limitation 
---at least at the present time--- for the mKdV reduction specified in Eq.~\eqref{asympototic results}.

\subsection{Dark breathers on the snoidal-wave background}\label{subsec: Dark breathers on snoidal waves}

We start the relevant presentation of waves on top of periodic
backgrounds by exploring the dark breather structure located on the 
sn-wave background of the FPUT lattice. Specifically, when the mKdV reduction in Eq.~\eqref{asympototic results} is defocusing (i.e., $K_4 < 0$, $K_2 > 0$), the dark-breather solution on the sn-wave background has been explicitly derived in Ref.~\cite{mucalica2024dark}. To formally introduce this exact solution, we need to introduce some necessary tools. Firstly, we introduce the first and fourth Jacobi's theta functions:
\begin{equation}\label{Jacobi's theta funcs}
    \begin{aligned}
    &\theta_1(y) = 2\sum_{n=1}^{\infty}(-1)^{n-1}q^{(n-\frac{1}{2})^2}\sin\left[(2n-1)y\right],\\
    &\theta_4(y) = 1 + 2\sum_{n=1}^{\infty}(-1)^nq^{n^2}\cos(2ny),
    \end{aligned}
\end{equation}
with $q = \exp\left(-\frac{\pi K(\sqrt{1-k^2})}{K(k)}\right)$, where $K(k)$ refers to the complete elliptic integral of the first kind.

Then, we further introduce the following functions:
\begin{equation}
    H(x) = \theta_1\left(\frac{\pi x}{2K(k)}\right),\quad \Theta(x) = \theta_4\left(\frac{\pi x}{2K(k)}\right),
\end{equation}
and the Jacobi's zeta function $Z(x) = \Theta'(x) / \Theta(x)$.

With these preliminaries, the exact solution reads \cite{mucalica2024dark}:
\begin{equation}\label{dark breather on snoidal-wave}
   \begin{aligned}
    &Y_{\text{db-sn}}({X},{T}) = \\
    &\sqrt{k}\frac{H(\xi+2\alpha)e^{-2\eta}+H(\xi-2\alpha)e^{2\eta}+2\beta H(\xi)}{\Theta(\xi+2\alpha)e^{-2\eta}+\Theta(\xi-2\alpha)e^{2\eta}+2\gamma\Theta(\xi)},
    \end{aligned}
\end{equation}
where $\xi = aX + c_0bT$ and $\eta = \kappa\left(aX+cb{T}+X_0\right)$ with
\begin{equation}\label{coefficients of a and b}
    \begin{aligned}
        a = \sqrt{6}\tilde{K_4}^{1/2}K_2^{-1/2}, \quad b = \frac{\sqrt{3}\tilde{K_4}^{3/2}}{2\sqrt{2}K_2^{1/2}c_s},
    \end{aligned}
\end{equation}
and $\tilde{K_4} = -K_4$. In addition, $X_0 \in \mathbb{R}$ is an arbitrary phase parameter, and the definitions all the other relevant parameters in Eq.~\eqref{dark breather on snoidal-wave} including $\beta,\gamma,\kappa,c_0,c$ are given in Appendix A.

\begin{figure*}[t!]
    \centering
    \includegraphics[width=0.325\linewidth]{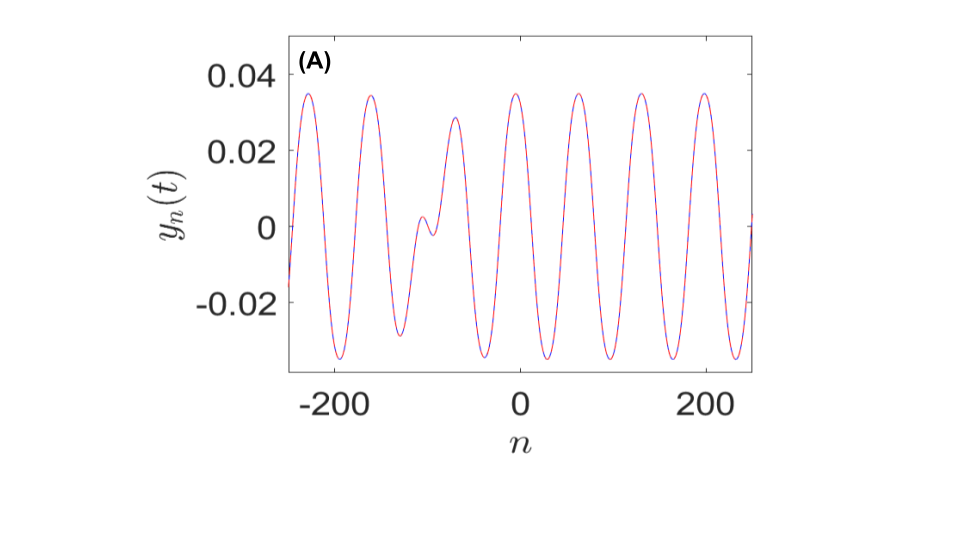}
    \hfill
    \includegraphics[width=0.325\linewidth]{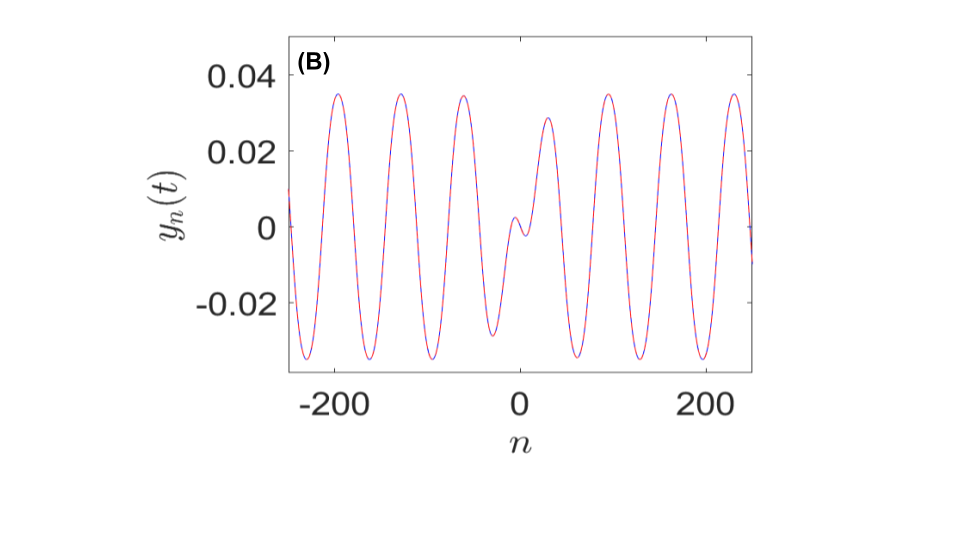}
    \hfill
    \includegraphics[width=0.325\linewidth]{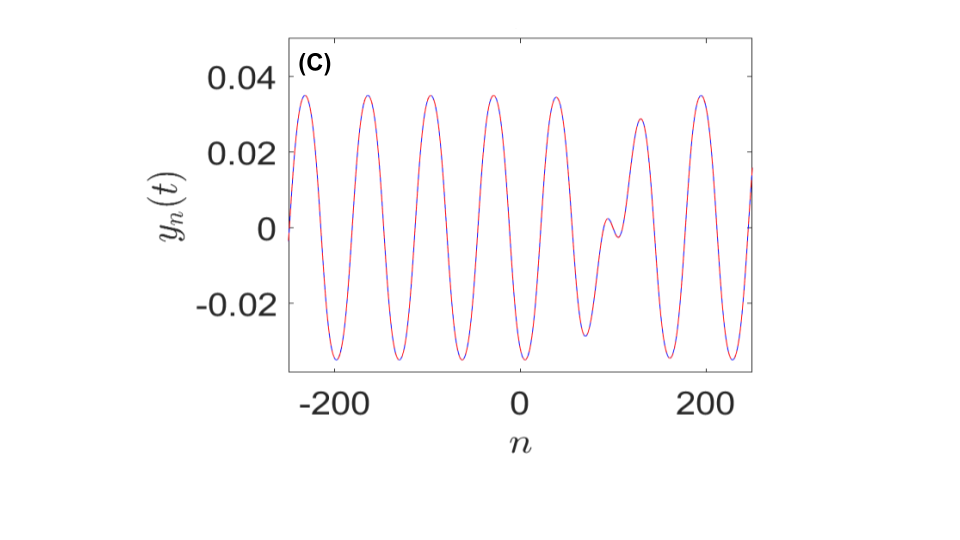}
    \caption{The comparison of the dark breather solution on the snoidal background. Panels (A), (B), (C) depict the comparisons made at $t = -100, 0, 100$, respectively, where $\epsilon = 0.05$, $k = 0.7$, and $K_2 = -K_4 = 1$. Finally, note that the solid blue and dashed red curves refer to the numerical solutions of the FPUT lattice \eqref{FPUT in strain} and the associated analytical solutions specified in Eq.~\eqref{dark breather on snoidal-wave}.}
    \label{fig:dark-breather on dn-background}
\end{figure*}

\begin{figure}[b!]
    \centering
    \includegraphics[width=0.5\linewidth]{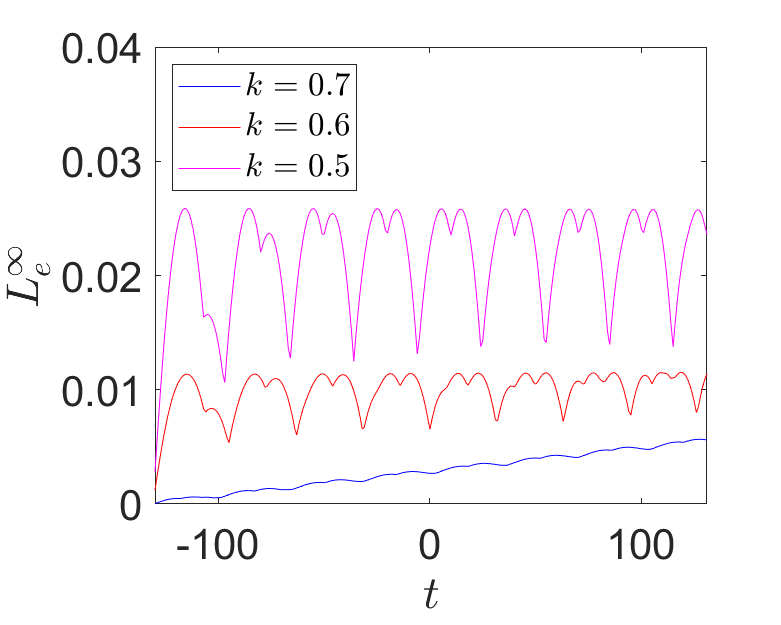}
    \caption{The $L^{\infty}_{e}$ losses for the dark-breather solution placed on the snoidal-wave background \eqref{dark breather on snoidal-wave}, for different
    values of the elliptic modulus, i.e., for $k=0.5$, $0.6$ and $0.7$. Note that $\epsilon = 0.05$.}
    \label{fig:DB error plot}
\end{figure}

Fig.~\ref{fig:dark-breather on dn-background} showcases the comparisons of the dark-breather wave profiles based on the FPUT dynamics and the associated analytical expression in Eq.~\eqref{dark breather on snoidal-wave} at $t = -100,0, 100$. We can clearly see that the numerical solutions of the FPUT lattice closely align with the exact solutions, suggesting a good agreement over the
space- and time-scales considered. We note that these good fits can be potentially explained by the modulationally stable feature of the 
sn-wave traveling-periodic background. {In addition, we have numerically explored other cases of $k$ and computed the associated relative $L^{\infty}$ error \eqref{eq: def of L_inf loss} at distinct time snapshots of $t$. Fig.~\ref{fig:DB error plot} shows the error plot where we can infer that the agreement between the analytical solutions in Eq.~\eqref{dark breather on snoidal-wave} and the corresponding numerical solutions of the FPUT lattice \eqref{FPUT in strain} is better as the value of the elliptic modulus $k$ increases}. To further corroborate this claim, we shall also explore the soliton placed on the modulationally stable dnoidal-wave background in Sec.~\ref{subsec: dn-wave for soliton}.

\subsection{Solitons on the dn-wave background}\label{subsec: dn-wave for soliton}

In the case of the focusing mKdV reduction in Eq.~\eqref{asympototic results} (i.e. $K_2, K_4>0$), the model naturally possesses the following soliton solution on the periodic-traveling dn-background \cite{chen2018rogue}, obtained through a nonlinear superposition of the Jacobi dn wave with an algebraically decaying soliton:
\begin{equation}\label{alg-decaying solitons on dn-wave background}
    Y_{\text{dn-alg}}(X,T) = Y_{\text{dn}} + \frac{\left(1-\theta_{\text{dn}}^2\right)\left(Y_{\text{dn}}^2+\sqrt{1-k^2}\right)}{\left(1+\theta_{\text{dn}}^2\right)Y_{\text{dn}}-\lambda_1^{-1}\theta_{\text{dn}}Y'_{\text{dn}}},
\end{equation}
with $\lambda_1 = \frac{1}{2}\left(1+\sqrt{1-k^2}\right)$, where
\begin{equation}
    \begin{aligned}
        &Y_{\text{dn}}(X,T) = \text{dn}\left(aX-cbT;k\right),\\
        &a = \frac{\sqrt{6}K_4^{1/2}}{K_2^{1/2}}, b = \frac{\sqrt{3}K_4^{3/2}}{2\sqrt{2}c_sK_2^{1/2}},c = 2-k^2,
    \end{aligned}
\end{equation}
and
\begin{equation}\label{def. of theta}
   \begin{aligned}
    &\theta_{\text{dn}}\left(X,T\right) =-4\lambda_1\left(Y_{\text{dn}}^2 + \sqrt{1-k^2}\right)\times \\
    &\left[\int_0^{aX-bcT}\frac{Y_{\text{dn}}(y)^2}{\left(Y_{\text{dn}}(y)^2 + \sqrt{1-k^2}\right)^2} - bT\right].
    \end{aligned}
\end{equation}

\begin{figure*}
    \centering
    \includegraphics[width=0.325\linewidth]{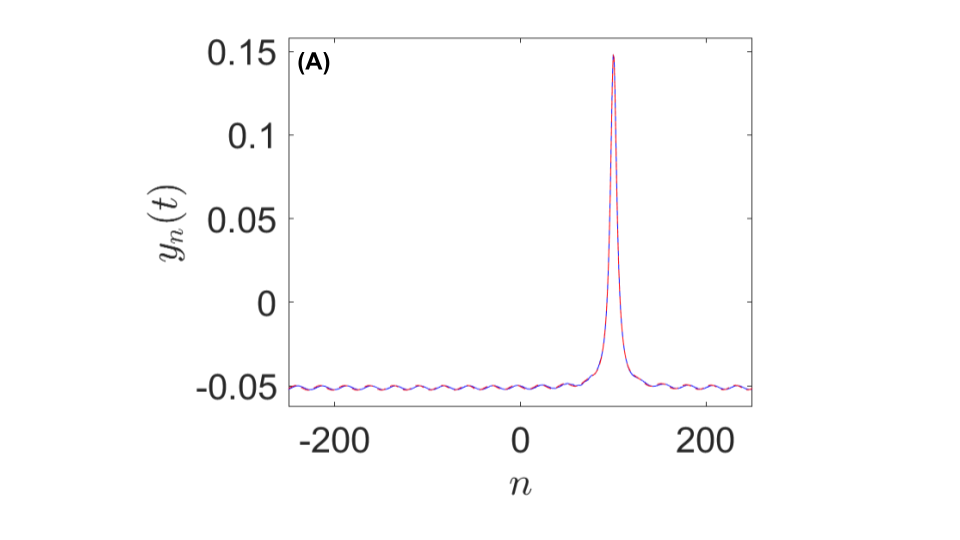}
    \hfill
    \includegraphics[width=0.325\linewidth]{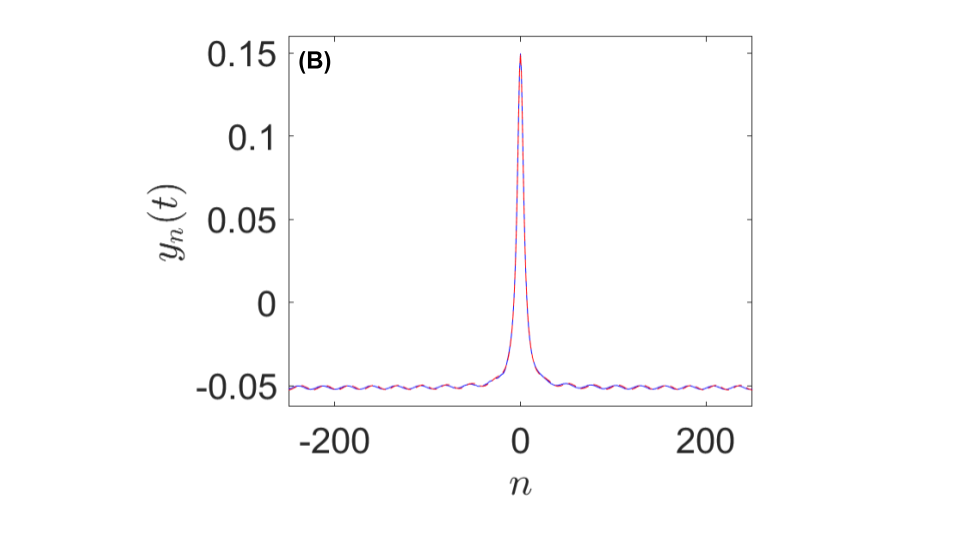}
    \hfill
    \includegraphics[width=0.325\linewidth]{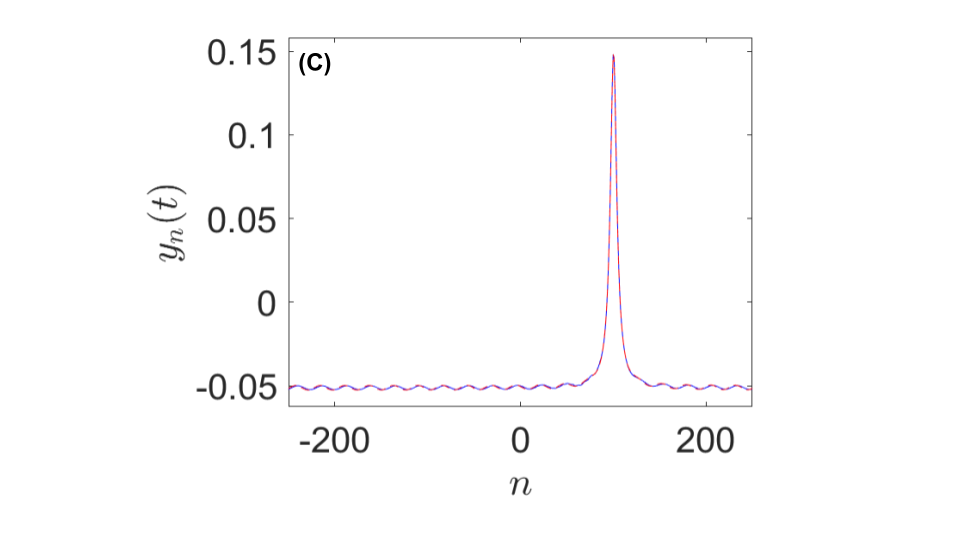}
    \caption{The comparison of the soliton solution on the dnoidal background. Panels (A), (B), (C) depict the comparisons made at $t = -100, 0, 100$, respectively, where $\epsilon = 0.05$, $k = 0.7$, and $K_2 = -K_4 = 1$. Finally, note that the solid blue and dashed red curves refer to the numerical solutions of the FPUT lattice \eqref{FPUT in strain} and the associated analytical solutions specified in Eq.~\eqref{alg-decaying solitons on dn-wave background}.}
    \label{fig:soliton solution on the dn-wave background}
\end{figure*}

Fig.~\ref{fig:soliton solution on the dn-wave background} depicts the relevant comparisons of the localized algebraically decaying soliton on the dnoidal wave background. Similar to the comparisons of the dark breather structure in Subsection~\ref{subsec: Dark breathers on snoidal waves}, the soliton structure of the numerical dynamics of the FPUT lattice fits very well with the analytical solutions
and appears to be naturally propagating through the FPUT lattice. Clearly, we shall expect such good agreement due to the modulation stability characteristics of the dnoidal waves \cite{chen2018rogue}.

\subsection{Rogue waves on the cnoidal-wave background}

We finally explore the exact solution of the mKdV equation in Eq.~\eqref{asympototic results} which represents a rogue wave on the cnoidal-wave background. Firstly, such solution takes the following form \cite{chen2018rogue}, which is obtained by applying the two-fold Darboux transformations:
\begin{equation}\label{rogue-wave on cn-wave}
    Y_{\text{cn-rogue}}(X,T) = Y_{\text{cn}} + \frac{G_1}{G_2},
\end{equation}
where
\begin{equation}
    \begin{aligned}
        &Y_{\text{cn}} = k\text{cn}\left(aX-bcT;k\right),\\
        &G_1 = 4k\sqrt{1-k^2}\text{Im}\bigg[\left(Y_{\text{cn}}^2+ik\sqrt{1-k^2}\right)(1-\theta_{\text{cn}}^2)\times\\
        &\left[(1+\overline{\theta}_{\text{cn}}^2)Y_{\text{cn}}-\overline{\lambda_I}^{-1}\overline{\theta}_{\text{cn}}Y'_{\text{cn}}\right]\bigg],\\
        &G_2 = (1-2k^2)\left|(1+\theta_{\text{cn}}^2)Y_{\text{cn}}-\lambda_I^{-1}\theta_{\text{cn}}Y'_{\text{cn}}\right|^2 \\
        &+ \left|1-\theta_{\text{cn}}^2\right|^2\left[Y_{\text{cn}}^4+k^2(1-k^2)\right] \\
        &+\left|(2\lambda_I)^{-1}(1+\theta_{\text{cn}}^2)Y_{\text{cn}}'-2\theta_{\text{cn}}Y_{\text{cn}}\right|^2,\\
        &c = 2k^2 - 1, \quad \lambda_I = \frac{1}{2}\left(k+i\sqrt{1-k^2}\right),
    \end{aligned}
\end{equation}
and
\begin{equation}\label{defn. of theta_cn}
   \begin{aligned}
    &\theta_{\text{cn}} = -4\lambda_I\left(Y_{\text{cn}}^2+ik\sqrt{1-k^2}\right) \times \\
    &\left[\int_0^{aX-bcT}\frac{Y_{cn}(y)^2}{\left(Y_{\text{cn}}(y)^2+ik\sqrt{1-k^2}\right)^2}dy - bT\right].
    \end{aligned}
\end{equation}
We observe that this rogue-wave solution in Eq.~\eqref{rogue-wave on cn-wave} obtains its global maximum at $(X,T) = (0,0)$, which resembles the rational solutions specified in Sec.~\ref{Sec: Local waves on constant back} except that the rogue wave now is superposed on the top of the cnoidal periodic-traveling wave background.

\begin{figure*}
    \centering
    \includegraphics[width=0.325\linewidth]{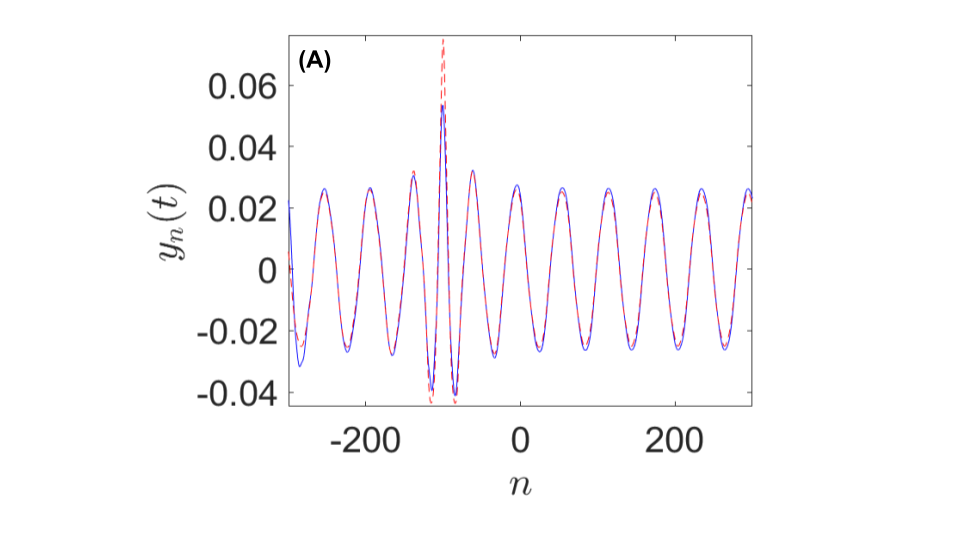}
    \hfill
    \includegraphics[width=0.325\linewidth]{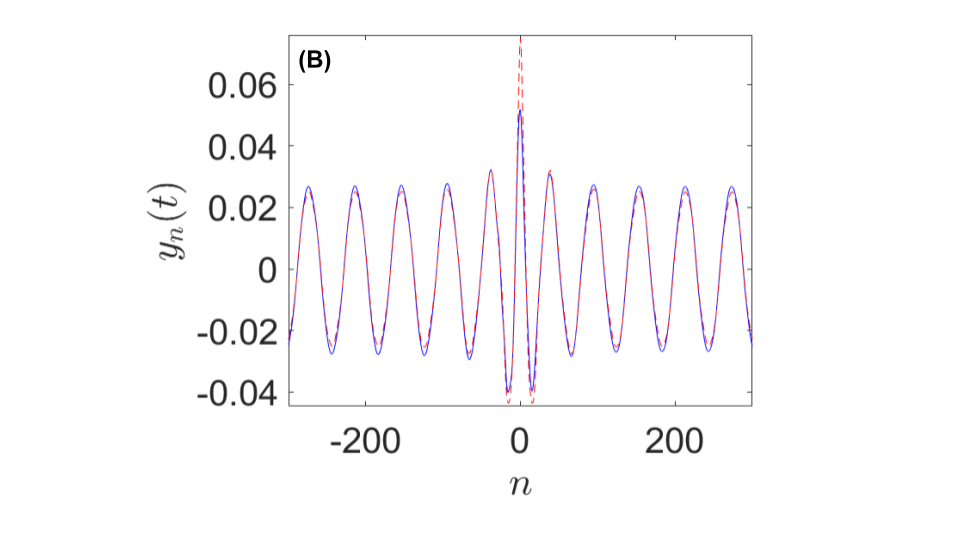}
    \hfill
    \includegraphics[width=0.325\linewidth]{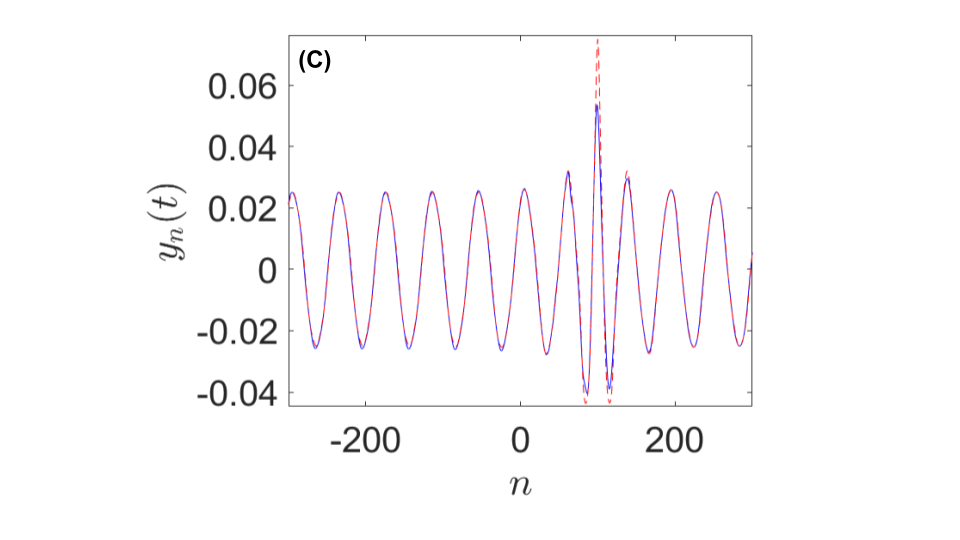}
    \caption{The comparison of the rogue-wave on the cnoidal-wave background. Notice that the values of these relevant parameters are $\epsilon = 0.05, k = 0.5, K_2 = K_4 = 1$. The numerical dynamics of the FPUT lattice in Eq.~\eqref{FPUT in strain} is similarly compared with its corresponding analytical counterparts \eqref{rogue-wave on cn-wave} at $t = -100, 0, 100$, respectively.}
    \label{fig:rogue-wave on cn-background comparison}
\end{figure*}

Fig.~\ref{fig:rogue-wave on cn-background comparison} shows the comparison of the numerical dynamics of the rogue waves on the cnoidal wave background. We notice that the relevant comparison demonstrates 
a deviation of the maximum wave amplitudes. Namely, the numerical dynamics always admits a maximum wave amplitude that is lower than that of the analytical solutions. These deviations are similar to those for the comparisons of the rational solutions specified in Sec.~\ref{Sec: Local waves on constant back}, and they can potentially result from the modulational instability feature of the cnoidal waves.
Nevertheless, we can observe a clear qualitative analogy between
the lattice numerical and the quasi-continuum analytical (approximate)
solutions which suggests that relevant waveforms are supported
by the FPUT lattice.

\section{Conclusions and future challenges}\label{Sec: conclusions and open directions}

In the present work, we have derived a modified KdV reduction associated with the FPUT lattice, and utilized its various analytical solutions which represent localized wave patterns to construct appropriate initial conditions for the FPUT lattice on top of a 
finite background, distinguishing our findings from earlier
work on regular solitonic excitations and their standard
parallels between the two models~\cite{VAINCHTEIN2022}. We have performed time stepping of these initial conditions and compared their evolutional dynamics with the associated analytical counterparts. The relevant comparisons for different localized wave structures have been shown to be reasonably good, which suggests the relevance of such analytical waveforms of the modified KdV reduction toward approximating the corresponding wave patterns (solitonic, but also
rogue, in homogeneous, but also in periodic backgrounds) 
within the FPUT lattice \eqref{FPUT in strain}. In most cases, 
we have found good quantitative agreement between the 
quasi-continuum theory and the numerical lattice evolution.
However, even when that is not the case (e.g. for rogue
waves in cnoidal backgrounds), there is a very good qualitative
match which suggests the relevance of the considered reductions.

We believe that this work paves the way for future research on approximating wave phenomena especially in the realm of
rogue waves and/or structures on top of a finite background
as they may emerge in the discrete lattice dynamics 
of the FPUT model. This has been achieved
by utilizing exact solutions from  quasi-continuum and long-wave models to approximate corresponding wave phenomena in discrete dynamical systems.
Along this vein, there are many open questions related to the present work. For example, one possible direction is to investigate higher-order rational solutions of the mKdV reduction with larger maximum wave amplitudes. In particular, it is also interesting to examine the performance of these higher-order rational solutions of the mKdV reduction in approximating the associated wave patterns of the FPUT lattice \eqref{FPUT in strain}. Secondly, it is also intriguing to consider the dimer FPUT lattice \cite{PhysRevE.98.032903} and similarly derive its corresponding mKdV reduction to study these distinct types of localized waves. Furthermore, one can also consider the two-dimensional FPUT lattice which is shown to admit the Kadomtsev–Petviashvili (KP) reduction \cite{hristov2022justification}. In particular, the KPI equation admits exact rogue-wave solutions \cite{XU201434,Wen_2017,ZHANG20181938}. Hence, leveraging an approach 
similar to the present work, one can establish the appropriate initial condition based on the analytical solutions of the KP equation for the two-dimensional FPUT lattice and henceforth explore their evolution dynamics. All of these perspectives lie in the realm of 
asymptotic multiscale reductions and associated computations.
However, it does not escape us that the path of such
connections between KdV-type models and FPUT-type lattices
is not only usable at the asymptotic level, but also at
the {\it rigorous} one, enabling one to prove the features
(existence, stability, etc.) of solitons within FPUT,
through connections to solitons in KdV. This path has been
followed for solitons~\cite{pegogf1,pegof2,pegof3,pegof4},
but also more recently for elliptic function (periodic)
solutions~\cite{Friesecke2015}. It would be of particular interest
to explore such ties in a similar way between $\beta$-FPUT and
mKdV.
All these directions are currently under consideration and will be reported in future publications.

\begin{acknowledgements}
    This material is based upon work supported by the National Science
Foundation under Grant No.
    PHY-2408988 (PGK). 
This research was partly conducted while P.G.K. was  visiting the Okinawa Institute of Science and
Technology (OIST) through the Theoretical Sciences Visiting Program (TSVP), the University of
Sydney through the visitor program of the Sydney Mathematical Research Institute (SMRI) and the Department of Mechanical Engineering at Seoul National 
University through a Fulbright Fellowship. Their support is gratefully acknowledged.
Finally, this work was also  supported by a grant from the Simons Foundation [SFI-MPS-SFM-00011048, P.G.K]. 

\end{acknowledgements}

\section*{Appendix A}\label{sec: Appendix A}

In this appendix A, we list the definitions of these parameters utilized to construct the analytical solutions for the dark breathers placed on the snoidal-wave background, as specified in Sec.~\ref{subsec: Dark breathers on snoidal waves}. In particular, these parameters of $\beta, \sigma, \kappa, c_0, c$ are defined as follows:
\begin{equation}\label{Defn of five para}
    \begin{aligned}
        &\beta = \left[1 - \frac{2(1+k)^2\text{sn}^2(\alpha)}{\left(1+k\text{sn}^2(\alpha)\right)^2}\right]\frac{\Theta(2\alpha)}{\Theta(0)}, \quad \gamma = \frac{\Theta(2\alpha)}{\Theta(0)},\\
        &\kappa = Z(\alpha) + \frac{k\text{sn}(\alpha)\text{cn}(\alpha)\text{dn}(\alpha)}{1 + k\text{sn}^2(\alpha)}, \quad c_0 = 1 + k^2,\\
        &c = c_0  \\ 
        &+\frac{2k(1+k)^2\left[1-k\text{sn}^2(\alpha)\right]\text{sn}(\alpha)\text{dn}(\alpha)\text{cn}(\alpha)}{\left[Z(\alpha)\left[1+k\text{sn}^2(\alpha)\right]+k\text{sn}(\alpha)\text{dn}(\alpha)\text{cn}(\alpha)\right]\left[1+k\text{sn}^2(\alpha)\right]^2}.
    \end{aligned}
\end{equation}

\bibliography{main}

\bibliographystyle{abbrv}

\end{document}